\documentclass[11pt,a4paper]{article} 
\usepackage{jheppub}

\usepackage{amsmath, amssymb}
\usepackage{concmath, palatino}
\usepackage{mathrsfs}

\usepackage{graphicx,epsfig}
\usepackage{epic}
\usepackage{color}

\newcommand{\be}{\begin{equation}}
\newcommand{\ee}{\end{equation}}
\newcommand{\ba}{\begin{eqnarray}}
\newcommand{\ea}{\end{eqnarray}}

\newcommand{\ads}{$AdS_5\times S^5$\ }

\newcommand{\wt}{\widetilde}
\newcommand{\wh}{\widehat}

\def\XXint#1#2#3{{\setbox0=\hbox{$#1{#2#3}{\int}$}
     \vcenter{\hbox{$#2#3$}}\kern-.5\wd0}}

    \newcommand{\beq}{\begin{equation}}
    \newcommand{\eeq}{\end{equation}}
    \newcommand\beqa{\begin{eqnarray}}
    \newcommand\eeqa{\end{eqnarray}}

\newcommand{\CC}{{\mathcal C}} 
\newcommand{\sql}{{ \sqrt\lambda}}
\newcommand{\E}{{\mathcal E}}

\title{Resummation of  semiclassical  short folded string}

\author[a]{Matteo Beccaria } 
\author[b]{, Guido Macorini} 

\affiliation[a]{Dipartimento di Fisica, Universita' del Salento \& INFN, \\
                     Via Arnesano, 73100 Lecce, Italy} 
                     
\affiliation[b]{Niels Bohr International Academy and Discovery Center,  \\
		Blegdamsvej 17 DK-2100 Copenhagen, Denmark}
                    
\emailAdd{matteo.beccaria@le.infn.it}
\emailAdd{macorini@nbi.ku.dk}

\abstract{
We reconsider semiclassical quantization  of  folded string spinning in $AdS_3$ part of  $AdS_5 \times
S^5$ using integrability-based (algebraic curve) method. We focus on the ``short string'' (small spin $S$)
limit with the angular momentum $J$ in $S^{5}$ scaled down according to $\mathcal J = \rho\,\sqrt \mathcal S$
in terms of the  variables $\mathcal J = J/\sqrt\lambda$, $\mathcal S = S/\sqrt\lambda$.
The semiclassical string energy in this particular scaling limit admits the double expansion 
$E = \sum_{n=0}^{\infty}\sum_{p=0}^{\infty} (\sqrt\lambda)^{1-n}\,a_{n,p}(\rho)\,\mathcal S^{p+1/2}$. 
It behaves smoothly as $J\to 0$ and partially resums recent results by Gromov and Valatka. 
We explicitly compute various one-loop coefficients $a_{1,p}(\rho)$ by summing over the 
fluctuation frequencies for integrable perturbations around the classical solution. 
For the simple folded string, the result agrees with what could be derived exploiting
a recent conjecture of Basso. However, 
the method  can be extended to more general situations. As an example, we consider
the $m$-folded string where Basso's conjecture fails. For this classical solution, we present the exact values
of $a_{1,0}(\rho)$ and $a_{1,1}(\rho)$ for $m=2, 3, 4, 5$ and explain how to work out the general case.
 }

\arxivnumber{1234.5678}

\begin{document} \maketitle

\bigskip

\section{Introduction and results}

AdS/CFT duality 
\cite{Maldacena:1997re} predicts the equivalence between the spectrum of the planar $\mathcal
 N = 4$ supersymmetric gauge theory and the spectrum of free closed quantum superstring propagating 
 in  \ads. States are  labeled both in the gauge and string theory by the five conserved spins  
 $C = (S_{1,2},  J_{1,2,3})$ corresponding to the 
bosonic subgroup $SO(2,4)\times SO(6)$  
of the symmetry
group $PSU(2,2 | 4)$ as well as by higher hidden charges. 
AdS/CFT correspondence can be expressed as the general relation
\be
E_{\rm gauge}(\lambda, C) = E_{\rm string}(\sqrt\lambda, C)\ ,
\ee
where 
$\frac{\sqrt\lambda}{2\pi} =
 \frac{R^{2}}{2\pi\alpha'}$ is  the \ads string tension in terms of the 't Hooft coupling $\lambda$.

In the strong-coupling regime ($\lambda\gg 1$) , massive 
quantum string states probe a near-flat region of \ads
and thus should have $E \sim \sqrt[4]\lambda$  \cite{Gubser:1998bc}.
More generally, considerations based on solving the two dimensional marginality 
condition \cite{Tseytlin:2003ac}    
perturbatively  in ${1 \over  \sqrt\lambda} \ll 1 $ 
  for fixed charges  suggest \cite{Roiban:2009aa}
   that (up to 
 a possible shift of $E$ by a
constant)\footnote{Here we assume   that the 1-loop string correction does not contain 
``non-analytic'' terms, cf.  \cite{Roiban:2009aa}.
This will be   indeed so  in the cases discussed below  treated in the 
algebraic curve approach.}
\be\label{1}
E^2 = 2N \sql + b_0 +   { b_1 \over \sqrt \lambda} + { b_2 \over (\sqrt \lambda)^2 }  + ...
\ ,
\ee
where $N$ is the flat-space level. 
As was argued in \cite{Tirziu:2008fk,Roiban:2009aa, Roiban:2011fe},  
one  can  attempt to find    quantum string energies 
   by starting with the  semiclassical 
strings with fixed  parameters $\CC= { C \over \sql}$
and then take 
 the {\em short} string limit $\CC \to 0$. Indeed, for quantum strings 
 with {\em fixed} charges $C$ the limit  $\sqrt \lambda \gg 1$  implies 
  $\CC = {C \over \sql} 
 \to 0$. Assuming  commutativity of the limits, 
 that  suggests the possibility 
  to compute the subleading  terms   in the above  expansion 
 by using the semiclassical  string theory methods. 
 The semiclassical string expansion gives 
 \be
\label{eq:general-semiclassical}
E =\sqrt{\lambda}\,\E_0(\CC)+\E_1(\CC)+\frac
1{\sqrt{\lambda}}\,\E_{2}(\CC)+\ldots\ .  
\ee 
Replacing  $\CC$ by $C \over \sql$
 and
 re-expanding  in  large $\lambda$ for fixed $C$ 
 one should find     that $E$ takes the form  consistent with (\ref{1}):
 \be
 E= \sqrt[4]\lambda \Big(k_1 + {k_2 \over \sql}   +  {k_3 \over (\sql)^2} + .... \Big) 
  \ . \ee
  This {\em semiclassical}  approach was successfully applied to the case of the 
 {\em short} string states representing members of the Konishi multiplet
 \cite{Roiban:2009aa,Gromov:2011de, Roiban:2011fe,Beccaria:2011uz,Gromov:2011bz}, 
 matching the results of the weak-coupling TBA approach extrapolated to strong coupling
\cite{Gromov:2009zb, Frolov:2010wt}. Also, the structure of the one-loop semiclassical energy $\mathcal E_{1}$
is very rich and can be exploited to make non trivial  higher order predictions \cite{Basso:2011rs,Gromov:2011bz}.

\bigskip
In this paper, we reconsider the specific case of the folded string spinning in $AdS_{5}$ and rotating in 
$S^{5}$~\cite{Frolov:2002av}
with the aim of giving some additional information about the one-loop correction $\mathcal E_{1}$.
For the quantization of this classical solution, 
the only charges appearing in (\ref{eq:general-semiclassical}) are the spin $\mathcal S = \frac{S}{\sqrt\lambda}$ and the angular momentum 
$\mathcal J = \frac{J}{\sqrt\lambda}$.
The short string expansion of the classical term $\mathcal E_{0}$ can be derived easily starting from the results in \cite{Frolov:2002av}
and reads (at fixed $\mathcal J$)
\ba
\label{eq:FT}
&& \mathcal E_{0}(\mathcal S, \mathcal J) = \mathcal J+\frac{\sqrt{\mathcal J^{2}+1}}{\mathcal J}\,\mathcal S-\frac{\mathcal J^{2}+2}{4\,\mathcal J^{3}(\mathcal J^{2}+1)}\,\mathcal S^{2}+ 
 \frac{2\,\mathcal J^{6}+\mathcal J^{4}+4\,\mathcal J^{2}+4}{64\,\mathcal J^{5}}\,\mathcal S^{3}+\cdots.
\ea
Following the analysis of \cite{Gromov:2011bz}, it is interesting to re-expand all the coefficients of the 
powers of $\mathcal S$ at $\mathcal J\to 0$. We obtain 
\ba
\label{eq:exp0}
\mathcal E_{0}(\mathcal S, \mathcal J) &=& \mathcal J+\left(
\frac{1}{\mathcal J}+\frac{\mathcal J}{2}-\frac{\mathcal J^{3}}{8}+\frac{\mathcal J^{5}}{16}+\cdots
\right)\,\mathcal S + \left(
-\frac{1}{2\,\mathcal J^{3}}+\frac{1}{4\,\mathcal J}-\frac{\mathcal J}{4}+\frac{\mathcal J^{3}}{4}+\cdots
\right)\,\mathcal S^{2}+ \nonumber \\
&& +\left(
\frac{1}{2\,\mathcal J^{5}}-\frac{1}{8\,\mathcal J}+\frac{11}{32}\,\mathcal J
-\frac{155}{256}\,\mathcal J^{3}+\cdots\right)\,\mathcal S^{3}+\cdots
\ea
The main remark of this paper is that this expansion breaks down at $\mathcal J\to 0$, but can be partially resummed by considering the  
scaling limit $\mathcal S\to 0$ with fixed ratio
\be
\label{eq:BM}
\frac{\mathcal J}{\sqrt\mathcal S} = \rho.
\ee
Indeed, in the regime (\ref{eq:BM}) the most singular terms in (\ref{eq:exp0}) are 
all proportional to $\sqrt\mathcal S$ with inverse powers of $\rho$
labeling the contributions coming from different powers of $1/\mathcal J$. 
Similarly, the next-to-leading singular terms would be 
resummed by the $\mathcal S^{3/2}$ contribution and so on. From the exact results of  \cite{Frolov:2002av}
it is a straightforward exercise to derive the following expansion~\footnote{Notice that (\ref{eq:classical-resummed}) is not a trivial consequence of  (\ref{eq:FT}), but is instead 
a genuine (partial) resummation thereof.}
\be
\label{eq:classical-resummed}
\mathcal E_{0}(\mathcal S, \rho) = \sqrt{\rho^{2}+2}\,\sqrt\mathcal S+
\frac{2\,\rho^{2}+3}{4\,\sqrt{\rho^{2}+2}}\,\mathcal S^{3/2}-
\frac{4\,\rho^{6}+20\,\rho^{4}+34\,\rho^{2}+21}{32\,(\rho^{2}+2)^{3/2}}\,\mathcal S^{5/2}+\cdots .
\ee

Consistency  of (\ref{eq:classical-resummed}) with (\ref{eq:exp0}) can be checked by expanding at large $\rho$
the various coefficients
\ba
\sqrt{\rho^{2}+2} &=& \rho+\frac{1}{\rho}-\frac{1}{2\,\rho^{3}}+\frac{1}{2\,\rho^{5}}+\cdots, \\
\frac{2\,\rho^{2}+3}{4\,\sqrt{\rho^{2}+2}} &=& \frac{\rho}{2}+\frac{1}{4\,\rho}-\frac{1}{8\,\rho^{5}}+\cdots, \\
-\frac{4\,\rho^{6}+20\,\rho^{4}+34\,\rho^{2}+21}{32\,(\rho^{2}+2)^{3/2}} &=&
-\frac{\rho^{3}}{8}-\frac{\rho}{4}-\frac{1}{8\,\rho}+\frac{1}{32\,\rho^{3}}+\cdots,
\ea
and  clearly (\ref{eq:classical-resummed}) includes an infinite number of  terms in (\ref{eq:exp0})
with arbitrary high powers of $\mathcal S$. 

\bigskip
A very natural question is whether the limit (\ref{eq:BM}) is also able to resum the one-loop contribution 
$\mathcal E_{1}(\mathcal S, \mathcal J)$ in a similar way.
In the recent paper \cite{Gromov:2011bz}, Gromov and Valatka  evaluated  $\mathcal E_{1}$ in the double limit
$\mathcal S\to 0$ followed by $\mathcal J\to 0$. The computation is performed   working in the algebraic curve 
framework  developed  in 
 \cite{Gromov:2007aq,Gromov:2007ky,Gromov:2008ec}.
An important result of \cite{Gromov:2011bz} is the following structure of the result
\be
\label{eq:GV}
\mathcal E_{1}(\mathcal S, \mathcal J) = \sum_{n=1}^{\infty}\mathcal S^{n}\,\sum_{p=0}^{\infty}\frac{c_{n, p}}{\mathcal J^{2n-1-2p}},
\ee
where the coefficients $c_{n,p}$ are rational combinations of zeta numbers $\zeta(n)$
and are computed at a certain fixed order in the $\mathcal S$ expansion. 
In the scaling regime (\ref{eq:BM}), we then obtain 
an expansion in half-integer powers of $\mathcal S$
\be
\label{eq:general-oneloop}
\mathcal E_{1}(\mathcal S, \rho) = \sum_{p=0}^{\infty}a_{1, p}(\rho)\,\mathcal S^{p+1/2},
\ee
where the large $\rho$ expansion of $a_{1, p}(\rho)$ generates all the coefficients $c_{n,p}$ with varying $n$
according to the relation
\be
a_{1, p}(\rho) = \sum_{n=1}^{\infty} \frac{c_{n,p}}{\rho^{2n-1-2p}}.
\ee
In more general terms, we expect the same structure to hold for all contributions $\mathcal E_{n}$, {\em i.e.}
\be
\label{eq:half}
\mathcal E_{n}(\mathcal S, \rho) = \sum_{p=0}^{\infty}a_{n, p}(\rho)\,\mathcal S^{p+1/2}.
\ee
 The explicit resummation can be performed by exploiting the recent Basso's conjecture \cite{Basso:2011rs}. 
According to this conjecture, the squared energy admits the expansion (compatible with semiclassical calculations)
\ba
\label{eq:basso} 
E^{2} &=& J^{2}+\left(A_{1}\,\sqrt\lambda+A_{2}+\frac{A_{3}}{\sqrt\lambda}
+\cdots\right)\,S + \left(B_{1}+\frac{B_{2}}{\sqrt\lambda}+ \frac{B_{3}}{\lambda}+\cdots\right)\,S^{2}+ \\
&&+
\left(\frac{C_{1}}{\sqrt\lambda}+\frac{C_{2}}{\lambda}+\frac{C_{3}}{\lambda^{3/2}}
+\cdots\right)\,S^{3}+\cdots. \nonumber
\ea
where the following exact formula for the constants $A_{i}$ is claimed
\be
A_{1}\,\sqrt\lambda+A_{2}+\frac{A_{3}}{\sqrt\lambda}+\cdots = 2\,\sqrt\lambda\,Y_{J}(\sqrt\lambda), \qquad
Y_{J}(x) = \frac{d}{dx}\,\log I_{J}(x).
\ee
Expanding at large $\lambda$, we find the first values~\footnote{We write them explicitly in order to emphasize the fact that the constants $A_{i}$ are polynomials in $J^{2}$. In general, also the
constants $B_{i}$, $C_{i}$ etc, will be dependent on $J^{2}$. }
\be
\begin{array}{ccl}
A_{1} &=& 2, \\
A_{2} &=& -1, \\
A_{3} &=& J^{2}-\frac{1}{4}, \\
A_{4} &=& J^{2}-\frac{1}{4}, 
\end{array}\qquad
\begin{array}{ccl}
A_{5} &=& -\frac{1}{4}\,J^{4}+\frac{13}{8}\,J^{2}-\frac{25}{64}, \\
A_{6} &=& -J^{4}+\frac{7}{2}\,J^{2}-\frac{13}{16}, \\
A_{7} &=& \frac{J^6}{8}-\frac{115 J^4}{32}+\frac{1187 J^2}{128}-\frac{1073}{512}.
\end{array}
\ee
Also, we know that $B_{1} = \frac{3}{2}$ and $B_{2} = \frac{3}{8}-3\,\zeta(3)$  \cite{Gromov:2011bz} .
Setting in (\ref{eq:basso}) the  scaling relation (\ref{eq:BM}) and comparing  with (\ref{eq:general-semiclassical}) and  (\ref{eq:half}) and find immediately
the following results
\ba
a_{1,0}(\rho) &=& -\frac{1}{2\,\sqrt{\rho^{2}+2}}, \\
a_{1,1}(\rho) &=& \frac{8 \rho ^4+23 \rho ^2+12}{16 \left(\rho ^2+2\right)^{3/2}}-\frac{3 \zeta (3)}{2 \sqrt{\rho^{2}
+2}}, \\
a_{2,0}(\rho) &=& -\frac{\rho^{2}+3}{8\,(\rho^{2}+2)^{3/2}}, \\
a_{3,0}(\rho) &=& -\frac{2\rho^{4}+9\rho^{2}+11}{16(\rho^{2}+2)^{5/2}},
\ea
and so on. Notice also that the scaling (\ref{eq:BM}) can be continued to the 
related regime where $\mathcal J/\mathcal S = J/S$ is kept fixed as in \cite{Gromov:2011de}. This continuation
cannot clearly be done at the level of  (\ref{eq:GV}), at least without resorting to Basso's conjecture.

\bigskip
The proposed resummation is based on Basso's results. In order to be able to treat more general cases, like for instance
the $m$-folded string \cite{Dorey:2008vp} where the conjecture is not valid, 
we want to show how to compute the functions $a_{1,n}(\rho)$ 
from the one-loop algebraic curve calculation in the regime (\ref{eq:BM}). 
This computation is interesting in itself since it shows how to compute the short string
expansion of a non trivial (elliptic) semiclassical solution by explicitly summing over the 
frequencies~\footnote{See also \cite{Gromov:2008fy} for a similar approach in a simpler case.} .
This raises various technical
points (missing poles, summation shifts) that could be important for the extension of such analysis to other cases like
ABJM or rigid circular strings.
In particular, 
we shall derive from a direct computation the functions $a_{1,0}(\rho)$ and $a_{1,1}(\rho)$
finding perfect agreement with the above results. We shall also  provide the following conjecture for $a_{1,2}(\rho)$ 
\be
a_{1,2}(\rho) = \frac{64 c-\rho^{2}\,(42\,\rho^{4}+212\,\rho^{2}+335)}{64\,(\rho^{2}+2)^{5/2}}
+\frac{3\,(4\,\rho^{4}+15\,\rho^{2}+13)}{8\,(\rho^{2}+2)^{3/2}}\,\zeta(3)+\frac{15\,(\rho^{2}+1)}
{8\,\sqrt{\rho^{2}+2}}\,\zeta(5),
\ee
depending on a single unfixed coefficient $c$. 

\bigskip
In the case of the $m$-folded string, Gromov and Valatka recently computed the 
expansion (\ref{eq:GV}) for various values of the folding number $m$. Their results are not compatible with 
any trivial modification of Basso's conjecture whose extension still has to be found. Our method can treat
this classical solution with minor effort. In particular, we are able to perform the resummation 
leading to (\ref{eq:general-oneloop}) and obtain the following leading one loop results 
\ba
\label{eq:new}
a_{1,0}^{m=2}(\rho) &=&
\frac{16\,\rho^{2}+5}{6\,\sqrt{\rho^{2}+2}}+\frac{1}{3}\,\sqrt{\rho^{2}+8}-3\,\rho, \nonumber \\
a_{1,0}^{m=3}(\rho) &=& 
\frac{243 \rho ^2}{40 \sqrt{\rho ^2+2}}+\frac{\sqrt{\rho ^2+18}}{8}+\frac{2}{5} \sqrt{4 \rho ^2+18}+\frac{53}{20 \sqrt{\rho ^2+2}}-7 \rho, \\
a_{1,0}^{m=4}(\rho) &=& \frac{1048 \rho ^2}{105 \sqrt{\rho ^2+2}}+\frac{\sqrt{\rho ^2+8}}{3}+\frac{\sqrt{\rho ^2+32}}{15}+\frac{3}{7} \sqrt{9 \rho ^2+32}+\frac{1007}{210 \sqrt{\rho ^2+2}}-\frac{35 \rho }{3}, \nonumber \\
a_{1,0}^{m=5}(\rho) &=& \frac{14375 \rho ^2}{1008 \sqrt{\rho ^2+2}}+\frac{\sqrt{\rho ^2+50}}{24}+\frac{2}{21} \sqrt{4 \rho ^2+50}+\frac{3}{16} \sqrt{9 \rho ^2+50}+\frac{4}{9} \sqrt{16 \rho ^2+50}+\nonumber \\
&&+\frac{3623}{504 \sqrt{\rho
   ^2+2}}-\frac{101 \rho }{6}.\nonumber
\ea
In Section (\ref{sec:beyond}) we shall explain how to work out a general $m$ without difficulty and 
provide various additional higher order results.
The expressions (\ref{eq:new}) 
have increasing complexity as $m$ becomes larger and suggest that the above modification is indeed
non trivial. Hopefully, they will be a useful constraint in fixing any proposed generalization of Basso's conjecture.

\section{The folded string in the Algebraic Curve framework}

The general construction of the algebraic curve for the \ads superstring is discussed in 
\cite{Gromov:2007aq,Gromov:2008ec}. Here, we summarize the main results in the specific case
of the folded string under consideration.

\subsection{Classical data}

The monodromy matrix of the Lax connection for the integrable dynamics of the \ads superstring has eigenvalues
\be
\{e^{i\,\widehat p_{1} }, e^{i\,\widehat p_{2} }, e^{i\,\widehat p_{3} }, e^{i\,\widehat p_{4} } |
e^{i\,\widetilde p_{1} }, e^{i\,\widetilde p_{2} }, e^{i\,\widetilde p_{3} }, e^{i\,\widetilde p_{4} }\}.
\ee
The eigenvalues are roots of the characteristic polynomial and define an 8-sheeted Riemann surface. The classical
algebraic curve has macroscopic cuts connecting various pairs of sheets. They impose suitable discontinuities
on the quasi-momenta $\wh p_{1,2,3,4}, \wt p_{1,2,3,4}$. In addition, one has to take into account Virasoro constraints and asymptotic properties
that are fully discussed in \cite{Gromov:2007aq,Gromov:2008ec} and that we shall not repeat here.

For the folded string
the classical solution is associated with an algebraic curve with two symmetric cuts along the real axis
\be
(-b,-a) \cup (a,b),\qquad 1<a<b.
\ee
The branch points are function of the charges 
\be
(\mathcal S, \mathcal J, \mathcal E) = \frac{1}{\sqrt\lambda}\,(S, J, E),
\ee
according to 
\ba
\mathcal S &=& \frac{1}{2\,\pi}\frac{ab+1}{ab}\,\left[b\,\mathbb E\left(1-\frac{a^{2}}{b^{2}}\right)
-a\,\mathbb K\left(1-\frac{a^{2}}{b^{2}}\right)\right], \nonumber \\
\mathcal J &=& \frac{1}{\pi}\,\frac{1}{b}\,\sqrt{(a^{2}-1)(b^{2}-1)}\,\mathbb K\left(1-\frac{a^{2}}{b^{2}}\right). \\
\mathcal E &=& \frac{1}{2\,\pi}\frac{ab-1}{ab}\,\left[b\,\mathbb E\left(1-\frac{a^{2}}{b^{2}}\right)
+a\,\mathbb K\left(1-\frac{a^{2}}{b^{2}}\right)\right].\nonumber
\ea
The complete set of quasi-momenta associated with this curve can be found in \cite{Gromov:2011de}. We do not 
repeat their expressions and simply follow the notation of that paper.

\subsection{One-loop correction to the energy}

The one-loop correction is computed in terms of two ingredients to be determined for each physical
polarization $I = (\alpha, \beta)$ taking the possible $8+8$ bosonic and fermionic values
\ba
S^{5} &:& (\alpha, \beta) = (\widetilde 1, \widetilde 3), (\widetilde 1, \widetilde 4), (\widetilde 2, \widetilde 3), (\widetilde 2, \widetilde 4), \\
AdS_{5} &:& (\alpha, \beta) = (\widehat 1, \widehat 3), (\widehat 1, \widehat 4), (\widehat 2, \widehat 3), (\widehat 2, \widehat 4), \\
\mbox{Fermions} &:& (\alpha, \beta) =  (\widetilde 1, \widehat 3), (\widetilde 1, \widehat 4), (\widetilde 2, \widehat 3), (\widetilde 2, \widehat 4), \\
&& \phantom{(i,j) =}\  (\widehat 1, \widetilde 3), (\widehat 1, \widetilde 4), (\widehat 2, \widetilde 3), (\widehat 2, \widetilde 4).\nonumber
\ea
The first ingredient is the on-shell pole associated with mode number $n$. It is computed by solving the
equation
\be
\label{eq:pole}
p_{\alpha}(x_{n}^{(I)})-p_{\beta}(x_{n}^{(I)}) = 2\,n\,\pi.
\ee
The meaning of $x_{n}^{(I)}$ is that of an extra quantum pole to be added to the classical 
cut representing the continuum limit of a dense distribution of classical Bethe roots for the integrable
classical solution of the superstring equations of motion.

The second ingredient is the off-shell fluctuation energy $\Omega^{(I)}(x)$. This is a quantity that provides
the one-loop correction to the energy from polarization $(I)$ once $x$ is replaced by its on-shell value
\be
\label{eq:omega}
\omega_{n}^{(I)} = \Omega^{(I)}(x_{n}^{(I)}).
\ee
As discussed in full details in \cite{Gromov:2008ec}, the folded string has enough simmetries to allow 
all off-shell fluctuation energies to be 
written in terms of only the two basic frequencies 
\ba
\Omega^{(S)} &=& {\Omega^{\widetilde 2\,\widetilde 3}(x) = 
\Omega^{\widetilde 2\,\widetilde 4}(x) = 
\Omega^{\widetilde 1\,\widetilde 3}(x) = 
\Omega^{\widetilde 1\,\widetilde 4}(x)  } = \ \ 
\frac{2}{ab-1}\frac{\sqrt{a^{2}-1}\sqrt{b^{2}-1}}{x^{2}-1}, \\
\Omega^{(A)} &=& {\Omega^{\widehat 2\,\widehat 3}(x)} = 
\frac{2}{ab-1}\,\left(1-\frac{f(x)}{x^{2}-1}\right).
\ea
where
\be
f(x) = \sqrt{x-a}\,\sqrt{x+a}\,\sqrt{x-b}\,\sqrt{x+b}.
\ee
All other frequencies are given by the following expressions
\ba
\Omega^{(1)}(x) &=& {\Omega^{\widehat 1\,\widehat 4}(x) }= -\Omega_{A}\left(\frac{1}{x}\right)-2, \\
\Omega^{(2)}(x) &=& {\Omega^{\widehat 1\,\widehat 3}(x) =\Omega^{\widehat 2\,\widehat 4}(x) }
= \frac{1}{2}\,\Omega_{A}(x)-\frac{1}{2}\,\Omega_{A}\left(\frac{1}{x}\right)-1, \\
\Omega^{(3)}(x) &=& {\Omega^{\widehat 2\,\widetilde 3}(x) = \Omega^{\widehat 2\,\widetilde 4}(x) = 
\Omega^{\widehat 3\,\widetilde 1}(x) = \Omega^{\widehat 3\,\widetilde 2}(x)} = \frac{1}{2}\,\Omega_{A}(x)+\frac{1}{2}\,\Omega_{S}(x), \\
\Omega^{(4)}(x) &=& {\Omega^{\widehat 1\,\widetilde 3}(x) = \Omega^{\widehat 1\,\widetilde 4}(x) = 
\Omega^{\widehat 4\,\widetilde 1}(x) = \Omega^{\widehat 4\,\widetilde 2}(x) }= \frac{1}{2}\,\Omega_{S}(x)-\frac{1}{2}\,\Omega_{A}\left(\frac{1}{x}\right)-1.
\ea
The one-loop shift of the energy is given in terms of the $\omega^{(I)}_{n}$
by the expression
\be
\label{eq:one-loop-correction}
\mathcal E_{1} = \frac{1}{2}\,\sum_{n,I} (-1)^{F_{I}}\,\omega_{n}^{(I)},
\ee
where $F_{I}$ is 1 for bosonic and $-1$ for fermionic polarizations. In principle, this sum is ill-defined
and could require separate non trivial shifts in the various terms as discussed in \cite{Gromov:2007aq}.
These ambiguities can be bypassed when there is a definite BMN limit \cite{Berenstein:2002jq}
of the classical solution.
Amazingly, the sum can also be written as a contour integral as in \cite{Gromov:2011de} apparently
solving automatically these problems. Here, we shall stick to the above representation as a sum over 
frequencies in order to provide all the details for a would-be interesting comparison with world-sheet
calculations. This comparison is however beyond the scope of this work and shall not be addressed.

\section{On-shell frequencies for physical polarizations: Leading order}

The explicit calculation of frequencies requires the solution of (\ref{eq:pole}). We shall find it perturbatively
in the short string limit $\mathcal S\to 0$ in the regime (\ref{eq:BM}). This limit can be implemented by 
using the following 
parametrization of the branch points
\ba
a &=& 1+s t+\frac{s^2 t^2}{2}+\frac{s^3 \left(-4 t^3-4 t^2+t\right)}{16 t+16}-\frac{s^4 \left(t^2 \left(2 t^3+6 t^2+4 t-1\right)\right)}{16 (t+1)}+\cdots,\\
b &=&1+s (t+2)+s^2 \left(\frac{t^2}{2}+2 t+2\right)+\frac{s^3 \left(4 t^3+20 t^2+31 t+14\right)}{16 t+16}+
\nonumber \\
&& -\frac{s^4 \left((t+2)^2 \left(2 t^3+6 t^2+4 t+1\right)\right)}{16 (t+1)}+\cdots, \nonumber
\ea
where $s$ is the expansion parameter and $t$ is a real constant.
The expansion is built in order to have 
\be
\label{eq:parameters}
\mathcal S = \frac{1}{2}\,s^{2}, \qquad
\frac{\mathcal J}{\sqrt\mathcal S}  =  \rho = \sqrt{2\,t\,(t+2)}.
\ee
Also, we can compute the expansion of the classical energy
\be
\frac{\mathcal E}{\sqrt{2\,\mathcal S}} = t+1+\frac{4t^{2}+8t+3}{16(t+1)}\,s^{2}+\cdots
\ee
In particular, for $t\to 0$ we recover the well known expansion of the classical energy of the $J=0$ folded string
\be
\mathcal E = \sqrt{2\,\mathcal S}\left(1+\frac{3}{8}\,\mathcal S+\cdots\right). 
\ee

\bigskip
The detailed analysis of the on-shell frequencies leads to the following results. We report the leading order expression of the pole position for $x>1$ (there is $x\to -x$ symmetry) and the full $\mathcal O(s)$ contribution to the on-shell energy correction obtained by computing
for each polarization $I$ the quantity (\ref{eq:omega})~\footnote{Notice that it requires 
the $\mathcal O(s^{3})$ expansion of the pole that we do not write for brevity.}.

\bigskip
\noindent
$\bullet \quad \mathbf{(\widetilde 2\,\widetilde 3) = (\widetilde 2\,\widetilde 4) = (\widetilde 1\,\widetilde 3) = (\widetilde 1\,\widetilde 4)}$
\bigskip

\noindent
There is a pole for each $n\ge 1$. It reads
\ba
x_{n}^{(S)} &=& 1+\frac{\sqrt{t\,(t+2)}}{n}\,s+\cdots, \\
\omega_{n}^{(S)} &=& \frac{n}{s (t+1)}-\frac{\sqrt{t (t+2)}}{t+1}+\frac{s \left(n^2 \left(4 t^2+8 t+5\right)+8 t (t+1)^2 (t+2)\right)}{16 n (t+1)^3}+\cdots. \nonumber 
\ea

\bigskip
\noindent
$\bullet\quad\mathbf{(\widehat 2\,\widehat 3)}$
\bigskip

\noindent
For $n=1$ there are two solutions discussed in the next section. For $n\ge 2$ there is a pole that reads
\ba
x_{n}^{(A)} &=& 1+\frac{\sqrt{(n-1)^{2}\,t\,(t+2)+1}-t-1}{n\,(n-2)}\,s+\cdots, \\
\omega_{n}^{(A)} &=& \frac{n}{s (t+1)}-1+ \nonumber \\
&& \ s \left[\frac{n^3 (4 t (t+2)+5)+n^2 (4 t (t+2) (2 t (t+2)+1)-6)-8 n t (t+2) (t+1)^2-8 (t+1)^4}{16 (n-2) n (t+1)^3}
\right. \nonumber \\
&&\qquad \left. -\frac{(n-1) \sqrt{(n-1)^2 t (t+2)+1}}{2 (n-2) n}\right]+\cdots.\nonumber
\ea

\bigskip
\noindent
$\bullet\quad\mathbf{(\widehat 1\,\widehat 4)}$
\bigskip

\noindent
There is a pole for each $n\ge 1$. It reads
\ba
x_{n}^{(1)} &=& 1+\frac{\sqrt{(n+1)^{2}\,t\,(t+2)+1}+t+1}{n\,(n+2)}\,s+\cdots, \\
\omega_{n}^{(1)} &=& \frac{n}{s (t+1)}-1+s \left[\frac{(n+1) \sqrt{(n+1)^2 t (t+2)+1}}{2 n^2+4 n}+\right. \nonumber \\
&&\quad \left. \frac{1}{16} \left(-\frac{8 \left(n^2+n-1\right) t}{(n+2) n}+\frac{n}{(t+1)^3}+\frac{4 (n+1)}{t+1}+4
   \left(\frac{1}{n+2}+\frac{1}{n}-2\right)\right)\right]+\cdots. \nonumber
\ea

\bigskip
\noindent
$\bullet\quad\mathbf{(\widehat 1\,\widehat 3) = (\widehat 2\,\widehat 4)}$
\bigskip

\noindent
For large enough $t$ there is a pole for $n\ge 2$. It reads
\ba
x_{n}^{(2)} &=& 1+\frac{\sqrt{(n^{2}-1)\,t\,(t+2)-1}}{n\,\sqrt{n^{2}-1}}\,s+\cdots, \\
\omega_{n}^{(2)} &=& \frac{n}{s (t+1)}-1+\frac{s \left(n^2 \left(4 t^2+8 t+5\right)+8 (t+1)^4\right)}{16 n (t+1)^3}
+\cdots.\nonumber
\ea
In the next section, we shall discuss what happens as $\rho$ decreases and eventually goes to zero.

\bigskip
\noindent
$\bullet\quad\mathbf{(\widehat 2\,\widetilde 3) = (\widehat 2\,\widetilde 4) =  
(\widehat 3\,\widetilde 1) = (\widehat 3\,\widetilde 2)}$
\bigskip

\noindent
For large enough $t$ there is a pole for $n\ge 2$. It reads
\ba
&& x_{n}^{(3)} = 1+\frac{(2\,n-1)\,\sqrt{t\,(t+2)}-t-1}{2\,n\,(n-1)}\,s+\cdots, \\
&& \omega^{(3)}_{n} = \frac{n}{s (t+1)}+\left(-\frac{\sqrt{t (t+2)}}{2 (t+1)}-\frac{1}{2}\right)+
s \left(\frac{\left(-n^2+n-1\right) \sqrt{t (t+2)}}{4 (n-1) n}+\right. \nonumber \\
&& \left. \frac{1}{16 (n-1) n (t+1)^3}(n^3 (4 t (t+2)+5)+n^2 (2 t (t+2) (2 t (t+2)+1)-3)+ \right. \nonumber \\
&& \left. + 2 n (t+1)^2 (2 t
   (t+2)+1)-2 (t+1)^2 (2 t (t+2)+1)\phantom{\frac{xxx}{xxx}}\right)+\cdots. \nonumber
\ea
In the next section, we shall discuss what happens as $\rho$ decreases and eventually goes to zero.

\bigskip
\noindent
$\bullet\quad\mathbf{(\widehat 1\,\widetilde 3) = (\widehat 1\,\widetilde 4) = (\widehat 4\,\widetilde 1)
= (\widehat 4\,\widetilde 2)}$
\bigskip

\noindent
For large enough $t$ there is a pole for $n\ge 2$. It reads
\ba
&& x_{n}^{(4)} = 1+\frac{(2\,n+1)\,\sqrt{t\,(t+2)}+t+1}{2\,n\,(n+1)}\,s+\cdots, \\
&&\omega^{(4)}_{n} = \frac{n}{s (t+1)}+\left(-\frac{\sqrt{t (t+2)}}{2 (t+1)}-\frac{1}{2}\right)+ s \left(
\frac{\sqrt{t (t+2)}}{4 n (n+1)}+\frac{1}{4} \sqrt{t (t+2)}+ \right. \nonumber \\
&&\left. + \frac{1}{16 n (n+1)
   (t+1)^3}
(n^3 (4 t (t+2)+5)-n^2 (2 t (t+2) (2 t (t+2)+1)-3) \right. \nonumber\\
&&\left. \qquad +2 n (t+1)^2 (2 t (t+2)+1)+2 (t+1)^2 (2 t (t+2)+1))\phantom{\frac{xxx}{xxx}}\right)+\cdots.\nonumber
\ea
In the next section, we shall discuss what happens as $\rho$ decreases and eventually goes to zero.

\subsection{Missing modes}

As we have seen, the basic equation (\ref{eq:pole}) admits a solution for all $n$ except special values. We shall refer
to these special values as 
{\em missing modes}. In general, one can locate the missing modes by taking a near BMN limit that 
in our case means large $\rho$ (or, what is the same, large $t$). In this regime, there is only a fixed and small number of missing poles 
while $\mathcal S\to 0$. Thus, it is easy to identify them and compute their contribution. Once this is done, it is 
possible to take $\rho\to 0$. In this process, additional missing solutions to (\ref{eq:pole}) do appear. They move
through the cuts and end on unphysical polarization planes as explained in \cite{Gromov:2008ec} . Their positions become generally complex. 
Nevertheless, their contribution to the energy is continuous through the crossing. So, the trick is to compute
the full one-loop correction at sufficiently large $\rho$ and then continue the result to small $\rho$.
A detailed analysis of the $n>0$ missing modes at large $\rho$ shows that 
\begin{enumerate}

\item[a)] The bosonic polarization $(A)$ has missing mode $n=1$. It corresponds to the branch point. The contribution 
can be smoothly computed as $\omega_{A,1}$ altough the expansion of the pole is singular for $n=1$.
The same is obtained by computing $\Omega^{(A)}(a) = \Omega^{(A)}(b)$. Notice that the multiplicity of this pole is one since a small deformation removes the pole from one of the two cut endpoints.

\item[b)] The bosonic polarization $(2)$ has missing mode $n= 1$. It can be computed by analitic continuation on the 
unphysical $(\wh 1 \wh 2)$ sheet. 
If this is done, one has to take into account that 2+2 poles are missing (the physical and unphysical ones at $n=\pm 1$).
They appear on the unphysical $(\wh 1 \wh 2)$ sheet at fixed positions $x=0, \pm i, \infty$.
Nevertheless, again it is possible to smoothly compute $\omega^{(2)}_{1}$.

\item[c)] The fermionic polarization $(3)$ (associated with $(\wh 2 \wt 3)$ ) has missing mode $n= 1$. These missing modes ($n=1$ in pair with $n=-1$) can be found on the unphysical sheet $(\wh 3 \wt 3)$ at fixed positions $x=0, \infty$. Their contribution 
requires the analytic continuation $\underline\Omega^{(3)}$ of 
\be
\Omega^{(3)} = \frac{1}{2}\Omega^{(A)}+\frac{1}{2}\Omega^{(S)}.
\ee
This is done simply by changing the sign of $f(x)$ inside $\Omega^{(A)}$ since the missing poles have passed through 
the cut. The final result is the average of the contributions at the two points which are
\ba
\underline \Omega^{(3)}(0) &=& \frac{1}{s(t+1)}-\frac{\sqrt{t(t+2)}}{t+1}+s\,\frac{8t^{4}+32t^{3}+48t^{2}+32t+9}{16(t+1)^{3}}+\cdots, \\
\underline \Omega^{(3)}(\infty) &=& \frac{1}{s(t+1)}-1+s\,\frac{8t^{4}+32t^{3}+48t^{2}+32t+9}{16(t+1)^{3}}+\cdots.
\ea
\end{enumerate}

\section{Sum over frequencies and $\mathcal O(\sqrt\mathcal S)$ correction}

The one loop correction is
\ba
\label{eq:bulk}
\mathcal E_{1} &=& \sum_{n=2}^{\infty}(4\,\omega_{n}^{(S)}+\omega^{(1)}_{n}+2\,\omega_{n}^{(2)}
+\omega_{n}^{(A)}-4\,\omega_{n}^{(3)}-4\,\omega_{n}^{(4)})+ \\ 
&& +4\,\omega_{1}^{(S)}+\omega^{(1)}_{1}+2\,\omega_{1}^{(2)}+\omega_{1}^{(A)}-4\,\omega_{1}^{(4)}+\nonumber \\
&& -4\times\frac{1}{2}(\underline \Omega^{(3)}(0) +\underline \Omega^{(3)}(\infty)).\nonumber
\ea
Using the previous results, we see that the $\sim 1/s$ and constant term $\sim s^{0}$ cancel. 
Instead, the $\mathcal O(s)$ contribution is non trivial and reads (here, LO means that we have not computed $\mathcal O(s^{2})$ terms)
\ba
\frac{\mathcal E_{1}^{\rm LO}}{s} &=& \sum_{n=2}^{\infty}\left[
\frac{n^2+6 t^2+12 t+2}{(n-2) n (n+1) (n+2) (n-1) (t+1)}-\frac{(n-1) \sqrt{(n-1)^2 t^2+2 (n-1)^2 t+1}}{2 (n-2) n}+
\right. \nonumber \\
&&\left. +\frac{(n+1) \sqrt{(n+1)^2 t^2+2 (n+1)^2 t+1}}{2 n (n+2)}+\frac{2 \sqrt{t (t+2)}}{n (n+1) (n-1)}
\right] + \\
&&+ \frac{10 t^2+20 t+1}{12 (t+1)}-\frac{3}{2} \sqrt{t (t+2)}+\frac{1}{3} \sqrt{4 t (t+2)+1}+\mathcal O(s)
.\nonumber
\ea
As a first check of this result we can take the $t\to 0$ limit. Then, we find
\be
\frac{\mathcal E_{1}^{\rm LO}}{s} = -\sum_{n=2}^{\infty}\frac{n^2+n+1}{(n-1) n (n+1) (n+2)}+\frac{5}{12} = -\frac{2}{3}+\frac{5}{12} = -\frac{1}{4}.
\ee
in agreement with \cite{Gromov:2011de}. 

\bigskip
The evaluation of the above sum for generic $t$ is apparently hopeless. However, due to the telescopic property
of the most complicated terms of the summand (those with $n$ inside the square roots), we can obtain the exact sum after some simple manipulations~\footnote{In some details, we are dealing with a sum of the form 
\be
\sum_{n=a}^{\infty} (f_{n}-f_{n-a}) = 
\lim_{N\to\infty}\sum_{n=a}^{N} (f_{n}-f_{n-a}) = -\sum_{n=0}^{a-1}f_{n}+a\,f_{\infty},
\ee
that reduces to a finite sum of $a$ terms plus a boundary contribution.}.
The result is remarkably simple and reads
\be
\label{eq:simple}
\frac{\mathcal E_{1}^{\rm LO}}{s} = -\frac{1}{4\,(t+1)}.
\ee
In terms of the scaled variable $\mathcal S, \mathcal J$ and trading $t$ for $\rho$ using
the second equation in (\ref{eq:parameters})
we finally find
\be
\frac{\mathcal E_{1}^{\rm LO}}{\sqrt{\mathcal S}} = -\frac{1}{2\,\sqrt{\rho^{2}+2}}
\ee

\section{$\mathcal O(\mathcal S)$ correction}

In general, we can expand the various on-shell energies $\omega_{n}^{(I)}$ in powers of $s$
\be
\omega_{n}^{(I)} = \omega_{n, -1}^{(I)}\,\frac{1}{s} + \omega_{n, 0}^{(I)}+
\omega_{n, 1}^{(I)}\,s+\omega_{n, 2}^{(I)}\,s^{2}+\cdots.
\ee
The $\mathcal O(\mathcal S)$ correction is associated to the contributions $\omega_{n, 2}^{(I)}$. This correction is rather simple and
reads
\ba
\omega^{(S)}_{n, 2} &=& -\frac{\sqrt{t (t+2)} \left(4 t^2+8 t+5\right)}{16 (t+1)^3}, \\
\omega^{(A)}_{n, 2} &=& 0,\\
\omega^{(1)}_{n, 2} &=& 0,\\
\omega^{(2)}_{n, 2} &=& 0,\\
\omega^{(3)}_{n, 2} &=& -\frac{\sqrt{t (t+2)} \left(4 t^2+8 t+5\right)}{32 (t+1)^3},\\
\omega^{(4)}_{n, 2} &=& -\frac{\sqrt{t (t+2)} \left(4 t^2+8 t+5\right)}{32 (t+1)^3}.
\ea
Adding the missing mode contribution, the full sum vanishes. This means that the one-loop correction to the energy has no $\mathcal O(s^{2})$
term, i.e. no $\mathcal O(\mathcal S)$ term. Of course, this is consistent with the general 
expansion (\ref{eq:general-oneloop}).

\section{$\mathcal O(\mathcal S^{3/2})$ correction and a conjecture for the $\mathcal O(\mathcal S^{5/2})$
contribution}

The $\mathcal O(\mathcal S^{3/2})$ correction is associated with the $\omega_{n, 3}^{(I)}$ terms. They are rather involved and we shall
not report them explicitly. Repeating the same kind of analysis of the LO correction~\footnote{There is only one remarkable technical point. The $n=1$ bosonic missing mode with polarization $(\wh 2 \wh 3)$ cannot be computed
by taking $n\to 1$ in the general $n\ge 2$ expression. Instead, one has to analitically continue in the unphysical 
plane $(\wh 2 \wh 1)$ as explained.} 
we are able to resum them
and the final result is 
\be
\label{eq:rho-expansion}
\mathcal E_{1} = a_{1,0}(\rho)\,{\mathcal S}^{1/2}+a_{1,1}(\rho)\,{\mathcal S}^{3/2}+a_{1,2}(\rho)\,{\mathcal S}^{5/2}+\cdots
\ee
with 
\be
{
\begin{array}{ccl}
a_{1,0}(\rho) &=& \displaystyle -\frac{1}{2\,\sqrt{\rho^{2}+2}}, \\ \\
a_{1,1}(\rho) &=& \displaystyle \frac{8 \rho ^4+23 \rho ^2+12}{16 \left(\rho ^2+2\right)^{3/2}}-\frac{3 \zeta (3)}{2 \sqrt{\rho ^2+2}}.
\end{array}
}
\ee
Thus, we have recovered with an explicit calculations the results that follow upon using Basso's conjecture.

\subsection{Matching the Gromov-Valatka expansion}

As a further check, we can expand at large $\rho$ finding
\ba
a_{1,0}(\rho) &=&  = -\frac{1}{2 \rho }+\frac{1}{2 \rho ^3}-\frac{3}{4 \rho ^5}+\frac{5}{4 \rho ^7}-\frac{35}{16 \rho ^9}+\cdots, \\
a_{1,1}(\rho) &=& \frac{\rho }{2}+\frac{-\frac{3 \zeta (3)}{2}-\frac{1}{16}}{\rho }+\frac{\frac{3 \zeta (3)}{2}+\frac{3}{16}}{\rho ^3}+\frac{-\frac{9 \zeta (3)}{4}-\frac{7}{32}}{\rho ^5}+\frac{\frac{15 \zeta
   (3)}{4}+\frac{5}{32}}{\rho ^7}+\frac{\frac{21}{128}-\frac{105 \zeta (3)}{16}}{\rho ^9}+
   \cdots\nonumber, 
\ea
in full agreement with (and extending !)  equation (B.5) of  \cite{Gromov:2011bz}. The remarkably simple
structure of these exact results suggest the following 
very reasonable conjecture for the function $a_{1,2}(\rho)$
\be
a_{1,2}(\rho) = \frac{\alpha_{0}+\alpha_{1}\,\rho^{2}+\alpha_{2}\,\rho^{4}+\alpha_{3}\,\rho^{6}}{(\rho^{2}+2)^{5/2}}
+\frac{\beta_{0}+\beta_{1}\,\rho^{2}+\beta_{2}\,\rho^{4}}{(\rho^{2}+2)^{3/2}}\,\zeta(3)+
\frac{\gamma_{0}+\gamma_{1}\,\rho^{2}}
{\sqrt{\rho^{2}+2}}\,\zeta(5).
\ee
Matching the Gromov-Valatka expansion we find immediately
\be
a_{1,2}(\rho) = \frac{64 c-\rho^{2}\,(42\,\rho^{4}+212\,\rho^{2}+335)}{64\,(\rho^{2}+2)^{5/2}}
+\frac{3\,(4\,\rho^{4}+15\,\rho^{2}+13)}{8\,(\rho^{2}+2)^{3/2}}\,\zeta(3)+\frac{15\,(\rho^{2}+1)}
{8\,\sqrt{\rho^{2}+2}}\,\zeta(5).
\ee
where $c$ is an undetermined rational constant. Of course, this is nothing but a conjecture and should be 
proved by an explicit computation at the needed order. This computation can be done along the lines
of the previous sections.

\subsection{The short string limit at fixed $r=J/S$ and the three-loop short state energy}

It is interesting to  remark that it is possible to set $\rho=r\,\sqrt\mathcal S$ in (\ref{eq:rho-expansion})
and discuss the short string limit
with fixed ratio $r = J/S$~\cite{Gromov:2011de}. The classical energy can be expanded at order $\mathcal S^{3}$ and is
\ba
\mathcal E_{0}(\mathcal S, r) &=& \sqrt{2\,\mathcal S}\,\left(
1+\frac{2r^{2}+3}{8}\,\mathcal S -\frac{4r^{4}-20r^{2}+21}{128}\,\mathcal S^{2}+\right. \\
&&\left. + 
\frac{8r^{6}-28r^{4}-146r^{2}+187}{1024}\,\mathcal S^{3}+
\cdots\right), \nonumber
\ea
while the one-loop result reads
\ba
\mathcal E_{1}(\mathcal S, r) &=& \sqrt{2\,\mathcal S}\,\left[
-\frac{1}{4}+\left(\frac{r^{2}+3}{16}-\frac{3}{4}\,\zeta(3)\right)\,\mathcal S+\right. \\
&&\left. + 
\left(\frac{-3r^{4}+28r^{2}+16c}{128}+\left(\frac{3r^{2}}{16}+\frac{39}{32}\right)\,\zeta(3)
+\frac{15}{16}\,\zeta(5)\right)\,\mathcal S^{2}+\cdots
\right].\nonumber
\ea
This expression gives some information about the three-loop strong coupling energy 
of short $(S, J)$ states. To this aim, we parametrize the higher order contributions $\mathcal E_{n}$  
\ba
\mathcal E_{n} &=& \sqrt{2\,\mathcal S}\left(\wt a_{n,0}(r)+\wt a_{n,1}(r)\,\mathcal S+\wt a_{n,2}(r)\,\mathcal S^{2}+\cdots\right).
\ea
Using the information we gathered on $\mathcal E_{0,1}$ the result is 
\ba
\frac{E}{\sqrt{2\,S}} &=& \lambda^{1/4}+\frac{1}{\lambda^{1/4}}\left(\frac{J^{2}}{4\,S}+\frac{3}{8}\,S
-\frac{1}{4}\right) + \\
&&
+\frac{1}{\lambda^{3/4}}\,\left(
-\frac{J^{4}}{32\,S^{2}}+\frac{J^{2}}{16\,S}+\frac{5J^{2}}{32}-\frac{21S^{2}}{128}-\frac{3\,\zeta(3)}{4}\,S
+\frac{3}{16}\,S+\wt a_{3,0}\right)+\nonumber \\
&&
+\frac{1}{\lambda^{5/4}}\left(
\wt a_{3,1}\,S+\wt a_{4,0}-\frac{3 J^4}{128 S^2}+J^2 \left(\frac{3 \zeta (3)}{16}+\frac{7}{32}\right)+S^2 \left(\frac{c}{8}+\frac{15 \zeta (5)}{16}+\frac{39 \zeta (3)}{32}\right)
\right)+\cdots\nonumber .
\ea
This expression can be compared with the two-loop prediction \cite{Gromov:2011bz} and matching 
is perfect upon setting $\wt a_{3,0} = -\frac{3}{32}$. Thus the three-loop predition contains the only 
undetermined  constant $c$
and the two functions $\wt a_{3,1}(r)$ and $\wt a_{4,0}(r)$.

\section{Beyond Basso's conjecture: The $m$-folded string}
\label{sec:beyond}

The considered folded string solution can be made more interesting by including the possibility of a  
higher folding, {\em i.e.} assuming that the string  bounces $m$ times
back and forth 
around the center of $AdS$
with a total of $2m$ spikes~\cite{Dorey:2008vp}. At the classical level, the modifications with respect to 
the simple $m=1$ case are trivial. In particular, we shall write again (\ref{eq:general-oneloop}) where now
\be
\mathcal S = \frac{S}{m\,\sqrt\lambda}, \qquad 
\mathcal J = \frac{J}{m\,\sqrt\lambda}.
\ee
The definitions of $s$, $\rho$, and $t$ are still given by (\ref{eq:BM}) and (\ref{eq:parameters}).
A naive attempt to modify in this way Basso's conjecture is known to be wrong already at 
one-loop \cite{Gromov:2011bz}.

In the Algebraic Curve approach, the higher folding $m$ is simply introduced by multiplying the 
charges and the quasi-momenta by $m$. Once this is done, the solutions of (\ref{eq:pole}) are obtained by 
replacing $n\to n/m$. This replacement does not affect the higher modes with $n\ge m$, 
but changes the low-lying ones. Let us consider in details the evaluation of $a_{1,0}^{m}(\rho)$ for the  first cases $m=2, 3, 4, 5$.

\bigskip\noindent
$\bullet\quad\mathbf m=2$
\bigskip

\noindent
The bulk contribution is obtained as the first line of (\ref{eq:bulk}) with $n\to n/2$ summed over $n\ge 3$.
The missing modes appear now at $n=2$ and their contribution is simply the contribution of the $n=1$
missing modes we computed for $m=1$. Finally, we have to add the genuine $n=1$ modes for the $m=2$ problem.
These are computed separately without problems (actually, all of them are obtained by setting $n=1/2$ in the $m=1$
expressions with the single exception of the $A$ polarization that has to be recomputed). Summing up, the final result is
\ba
\frac{\mathcal E_{1}^{\rm LO, m=2}}{s} &=& \sum_{n=3}^{\infty}\left[
\frac{(n+2) \sqrt{(n+2)^2 t^2+2 (n+2)^2 t+4}}{2 n^2+8 n}+\frac{8 \left(n^2+24 t^2+48 t+8\right)}{(n-4) n (n+2) (n+4) (n-2)
   (t+1)}+\right. \nonumber\\
   &&\left. + \frac{16 \sqrt{t (t+2)}}{n \left(n^2-4\right)}-\frac{(n-2) \sqrt{(n-2)^2 t^2+2 (n-2)^2 t+4}}{2 (n-4) n}
\right] +\frac{306 t^2+612 t+101}{60 t+60}+\nonumber \\
&&-\frac{41}{6} \sqrt{t (t+2)}+\frac{1}{6} \sqrt{t (t+2)+4}+\frac{1}{3} \sqrt{4 t
   (t+2)+1}+\frac{3}{10} \sqrt{9 t (t+2)+4}
\ea
Using again the telescopic property of the most complicated terms in the summand, we are able to compute the 
exact infinite sum and obtain
\be
\frac{\mathcal E_{1}^{\rm LO, m=2}}{s} = \frac{8}{3}(t+1)-\frac{9}{4\,(t+1)}-3\,\sqrt{t\,(t+2)}
+\frac{1}{3}\,\sqrt{t\,(t+2)+4}.
\ee
Replacing $t = \frac{1}{2}(\sqrt{2\rho^{2}+4}-2)$ we can write the exact result
\be
\label{eq:winding}
a_{1,0}^{m=2}(\rho) = 
\frac{16\,\rho^{2}+5}{6\,\sqrt{\rho^{2}+2}}+\frac{1}{3}\,\sqrt{\rho^{2}+8}-3\,\rho. 
\ee
Expanding at large $\rho$ we find 
\be
a_{1,0}^{m=2}(\rho) =
 -\frac{1}{2 \rho }+\frac{1}{2 \rho ^3}+\frac{21}{4 \rho ^5}-\frac{175}{4 \rho ^7}+\frac{4501}{16 \rho ^9}
 -\frac{28161}{16\,\rho^{11}}+\frac{358545}{32\,\rho^{13}}+\cdots,
 \ee
 in full agreement with (and extending) the leading singular terms of (B.6) of \cite{Gromov:2011bz}. 
  The same kind of computation can be repeated for the next higher order term in the $\mathcal S$ expansion with
 the rather simple result
 \ba
 a_{1,1}^{m=2}(\rho) &=& \frac{15}{4} \left(\rho ^3+\rho \right)-\frac{85 \rho ^4+721 \rho ^2+940}{108 \sqrt{\rho ^2+8}}-\frac{1280 \rho ^6+3720 \rho ^4+1635 \rho ^2-884}{432 \left(\rho ^2+2\right)^{3/2}}+\nonumber \\
 && -\frac{12 \zeta (3)}{\sqrt{\rho
   ^2+2}}.
 \ea
 Again, expanding at large $\rho$ we find 
 \be
 a_{1,1}^{m=2}(\rho) = \frac{\rho }{2}+\frac{-12 \zeta (3)-\frac{17}{16}}{\rho }+\frac{12 \zeta (3)+\frac{19}{16}}{\rho ^3}+\frac{-18 \zeta (3)-\frac{727}{32}}{\rho ^5}+\frac{30 \zeta (3)+\frac{8365}{32}}{\rho ^7}+\cdots,
 \ee
 in full agreement with (and extending) the next-to-leading singular terms of (B.6) of \cite{Gromov:2011bz}.

\bigskip
\noindent
$\bullet\quad\mathbf m=3$
\bigskip

\noindent 
 The same kind of manipulations leads to the result
 \be
\label{eq:winding}
a_{1,0}^{m=3}(\rho) = 
\frac{243 \rho ^2}{40 \sqrt{\rho ^2+2}}+\frac{\sqrt{\rho ^2+18}}{8}+\frac{2}{5} \sqrt{4 \rho ^2+18}+\frac{53}{20 \sqrt{\rho ^2+2}}-7 \rho.
\ee
Expanding at large $\rho$ we find 
\be
a_{1,0}^{m=3}(\rho) =
-\frac{1}{2 \rho }-\frac{5}{8 \rho ^3}+\frac{1245}{32 \rho ^5}-\frac{258785}{512 \rho ^7}+\frac{13235411}{2048 \rho ^9}+\cdots,
 \ee
 in full agreement with (and extending) the leading singular terms of (B.7) of \cite{Gromov:2011bz}. 
At the next order, we find a rather complicated closed expression for $a_{1,1}^{m=3}(\rho)$. Its 
expansion at large $\rho$ is
 \be
 a_{1,1}^{m=3}(\rho) =\frac{\rho }{2}+\frac{-\frac{81 \zeta (3)}{2}-\frac{7}{4}}{\rho }+\frac{\frac{81 \zeta (3)}{2}+\frac{39}{16}}{\rho ^3}+\frac{-\frac{243 \zeta (3)}{4}-\frac{251423}{1024}}{\rho ^5}+O\left(\left(\frac{1}{\rho
   }\right)^{11/2}\right)
 \ee
 in full agreement with the next-to-leading singular terms of (B.7) of \cite{Gromov:2011bz}. 

\bigskip
\noindent
$\bullet\quad\mathbf m=4$
\bigskip

\noindent 
The exact result is 
 \be
\label{eq:winding}
a_{1,0}^{m=4}(\rho) = \frac{1048 \rho ^2}{105 \sqrt{\rho ^2+2}}+\frac{\sqrt{\rho ^2+8}}{3}+\frac{\sqrt{\rho ^2+32}}{15}+\frac{3}{7} \sqrt{9 \rho ^2+32}+\frac{1007}{210 \sqrt{\rho ^2+2}}-\frac{35 \rho }{3}.
\ee
Expanding at large $\rho$ we find 
\be
a_{1,0}^{m=4}(\rho) =-\frac{1}{2 \rho }-\frac{55}{18 \rho ^3}+\frac{43109}{324 \rho ^5}-\frac{8049175}{2916 \rho ^7}+\frac{6448461173}{104976 \rho ^9}+\cdots.
 \ee
 At the next order, we find a complicated closed expression for $a_{1,1}^{m=4}(\rho)$ whose expansion 
 at large $\rho$ reads
 \be
 a_{1,1}^{m=4}(\rho) =\frac{\rho }{2}+\frac{-96 \zeta (3)-\frac{281}{144}}{\rho }+\frac{96 \zeta (3)+\frac{185}{48}}{\rho ^3}+\frac{-144 \zeta (3)-\frac{31420351}{23328}}{\rho ^5}+\cdots.
 \ee
 It implies that the analog of Eqs.~(B.5, B.6, B.7) of \cite{Gromov:2011bz} for $m=4$ should read (in the notation 
 of that paper)
 \ba
 \Delta_{\rm 1-loop}^{m=4} &=& \left(-\frac{1}{2\,\mathcal J}+\frac{\mathcal J}{2}+\cdots\right)\,\mathcal S + 
 \left[
 -\frac{55}{18\,\mathcal J^{3}}
 -\frac{1}{\mathcal J}\,\left(
 -96 \zeta (3)-\frac{281}{144}
 \right)
 +\cdots\right]\,\mathcal S^{2}+ \nonumber \\
&&+ \left[\frac{43109}{324\,\mathcal J^{5}}
 +\frac{1}{\mathcal J^{3}}\,\left(
 96 \zeta (3)+\frac{185}{48}
 \right)
 +\cdots\right]\,\mathcal S^{3}+\cdots
 \ea

\bigskip
\noindent
$\bullet\quad\mathbf m=5$
\bigskip

\noindent 
The exact result is 
 \ba
\label{eq:winding}
a_{1,0}^{m=5}(\rho) &=& \frac{14375 \rho ^2}{1008 \sqrt{\rho ^2+2}}+\frac{\sqrt{\rho ^2+50}}{24}+\frac{2}{21} \sqrt{4 \rho ^2+50}+\frac{3}{16} \sqrt{9 \rho ^2+50}+\frac{4}{9} \sqrt{16 \rho ^2+50}+\nonumber \\
&&+\frac{3623}{504 \sqrt{\rho
   ^2+2}}-\frac{101 \rho }{6}.
\ea
Expanding at large $\rho$ we find
\be
a_{1,0}^{m=5}(\rho) =-\frac{1}{2 \rho }-\frac{1981}{288 \rho ^3}+\frac{13823521}{41472 \rho ^5}-\frac{246936383285}{23887872 \rho ^7}+\frac{1230239204504695}{3439853568 \rho ^9}+\cdots .
 \ee
 At the next order, we compute $a_{1,1}^{m=5}(\rho)$ whose expansion at large $\rho$ is 
 \be
 a_{1,1}^{m=5}(\rho) =\frac{\rho }{2}+\frac{-\frac{375 \zeta (3)}{2}-\frac{911}{576}}{\rho }+\frac{\frac{375 \zeta (3)}{2}+\frac{259}{48}}{\rho ^3}+\frac{-\frac{1125 \zeta (3)}{4}-\frac{242572445069}{47775744}}{\rho
   ^5}+\cdots.
 \ee
 It implies that the analog of Eqs.~(B.5, B.6, B.7) of \cite{Gromov:2011bz} for $m=5$ reads (in the notation 
 of that paper)
 \ba
 \Delta_{\rm 1-loop}^{m=5} &=& \left(-\frac{1}{2\,\mathcal J}+\frac{\mathcal J}{2}+\cdots\right)\,\mathcal S + 
 \left[
-\frac{1981}{288\,\mathcal J^{3}}
 -\frac{1}{\mathcal J}\,\left(
-\frac{375 \zeta (3)}{2}-\frac{911}{576}
 \right)
 +\cdots\right]\,\mathcal S^{2}+ \nonumber \\
&&+ \left[\frac{13823521}{41472\,\mathcal J^{5}}
 +\frac{1}{\mathcal J^{3}}\,\left(
\frac{375 \zeta (3)}{2}+\frac{259}{48}
 \right)
 +\cdots\right]\,\mathcal S^{3}+\cdots
 \ea

\subsection{Resummed values at $J=0$}

It is interesting to compute the values of the $a(\rho)$ functions at $\rho=0$, which is $J=0$. They are a non trivial
result of the resummation. Using the closed expressions we have computed, one obtains immediately the following table~\footnote{Notice that $a_{1,1}^{m=1}(0)$ must not be identified with the similar
constant $B_{2}$. Instead, the precise relation from (\ref{eq:basso}) is $B_{2}=2\,\sqrt 2\,a_{1,1}^{m=1}(0)-3/8$.
Replacing the value of the table, we get back the result of Gromov and Valatka.}
\be
\begin{array}{||c|c|c||}
\hline
m & \sqrt2\,a_{1,0}^{m}(0) & \sqrt 2\,a_{1,1}^{m}(0) \\
\hline
&& \\
\ \ 1\ \  & \displaystyle -\frac{1}{2} & \displaystyle \frac{3}{8}-\frac{3}{2}\,\zeta_{3} \\
&& \\ \hline && \\
\ \ 2\ \  & \displaystyle \frac{13}{6} & \displaystyle -\frac{719}{216}-12\,\zeta_{3} \\
&& \\ \hline && \\
\ \ 3\ \ & \displaystyle \frac{29}{5} & \displaystyle -\frac{89547}{8000}-\frac{81}{2}\,\zeta_{3} \\
&& \\ \hline && \\
\ \ 4\ \ & \displaystyle \frac{2119}{210} & \displaystyle -\frac{230235893}{9261000}-96\,\zeta_{3} \\
&& \\ \hline && \\
\ \ 5\ \ & \displaystyle \frac{3749}{252} & \displaystyle -\frac{5894598851}{128024064}-\frac{375}{2}\,\zeta_{3} \\
&& \\ \hline
\end{array}
\ee
As a final check, we remark that the values $a_{1,0}^{m}(0)$ agree with a general analysis of the short  $m$-folded string in the fixed 
$J/S$ regime  \cite{prep} as they should.
  

\section*{Acknowledgments}

We thank Arkady Tseytlin for many interesting discussions and for pointing out the role of Basso's conjecture
in the resummation.
We also thank N. Gromov for helpful clarifications on the algebraic curve approach.

\appendix

\bibliography{AC-Biblio}{}
\bibliographystyle{JHEP}

\end{document}